\documentclass[groupedaddress,twocolumn,prl]{revtex4}
\usepackage{epsfig,amsfonts,amsmath,graphicx,subfigure}

\newcommand{\be}{\begin{equation}}
\newcommand{\ee}{\end{equation}}
\newcommand{\ba}{\begin{eqnarray}}
\newcommand{\ea}{\end{eqnarray}}

\def\bea{\begin{eqnarray}}
\def\eea{\end{eqnarray}}
\def\ben{\begin{eqnarray*}}
\def\een{\end{eqnarray*}}

\def\>{\rangle}
\def\<{\langle}
\def\l{\left}
\def\r{\right}

\newcommand{\fig}[1]{Fig.~\ref{fig:#1}}

\begin{document}

\title{Electron impact ionization loading of a surface electrode ion trap}

\author{Kenneth R. Brown, Robert J. Clark, Jaroslaw Labaziewicz,
Philip Richerme, David R. Leibrandt, and Isaac L. Chuang}

\affiliation{Center for Bits and Atoms, Research Laboratory of
Electronics, \& Department of Physics\\ Massachusetts Institute of
Technology, Cambridge, Massachusetts 02139, USA}

\date{\today}

\begin{abstract}

We demonstrate a method for loading surface electrode ion traps by electron impact ionization. The method relies on the property of surface electrode geometries that the trap depth can be increased at the cost of more micromotion. By introducing a buffer gas, we can counteract the rf heating assocated with the micromotion and benefit from the larger trap depth. After an initial loading of the trap, standard compensation techniques can be used to cancel the stray fields resulting from charged dielectric and allow for the loading of the trap at ultra-high vacuum.

\end{abstract}


\maketitle

%%%%%%%%%%%%%%%%%%%%%%%%%%%%%%%%%%%%%%%%%%%%%%%%%%%%%%%%%%%%%%%%%%%%
% Introduction 
%%%%%%%%%%%%%%%%%%%%%%%%%%%%%%%%%%%%%%%%%%%%%%%%%%%%%%%%%%%%%%%%%%%%

%Surface traps are kewl

Surface electrode ion traps \cite{Chiaverini:05, Pearson:06, Seidelin:06, Britton:06} offer significant potential for realizing complicated
geometries needed for large-scale quantum computation
\cite{Kielpinski:02}. Their advantages include greater ease of
fabrication than three dimensional (3D) layered planar traps
\cite{Barrett:04, Madsen:04, Monroe:06} and the ability to integrate control
electronics below the electrode surface \cite{Slusher:05}. However, the 2D geometery results in a shallow trap depth, which is $\sim 1/100$ of comparably
sized 3D traps \cite{Chiaverini:05}. In the presence of stray electric fields, the depth can become even shallower.

Stray electric fields can also displace ions from the zero point of the trap radiofrequency (rf) field. This causes undesired heating of ions, resulting from coupling of the rf-driven ``micromotion'' of one ion with the secular motion of neighboring ions. A well-developed technique to mitigate this effect is to apply dc \emph{compensation} voltages, usually to special electrodes placed around the ion \cite{Berkeland:98,Raab:00,Lisowski:05}. 

To find the experimental dc compensation values, one typically starts with the compensation values for an ideal trap. For a symmetric 3D linear trap, the expected compensation values are zero. The asymmetry of the 2D linear trap requires numerically solving Maxwell's equations to find compensated dc electrode values \cite{Seidelin:06} since the dc voltages used to confine the ions axially also shift the ion positions vertically, and since dielectric insulators needed between electrodes are neglected in analytical solutions. If an ion signal is easily observed, the experimental compensated values can be quickly found. However, large stray fields or trap imperfections often impede observation of ion signal and a random walk of the compensation voltages must be undertaken. 

Electron impact ionization, the standard method for loading ion traps, charges dielectrics in the vacuum chamber, leading to large stray fields.  Photoionization can be used to avoid creating stray charge at the cost of additional lasers and has been used to load shallow 2D and 3D traps \cite{Seidelin:06, Monroe:06}. Here we demonstrate a method for loading 2D traps with electron impact ionization that relies on the asymmetry of the trap and a buffer gas to obtain the initial signal.

An uncompensated trap leads to an increase in micromotion and is never advantageous for a 3D geometry. However, for a 2D geometry an applied field perpendicular to the surface can result in a significantly deeper trap in exchange for more micromotion \cite{Pearson:06}. In this setting, the number of ions loaded increases but laser cooling is not efficient enough to counter the rf heating, causing the ions to escape.

The increase in rf heating can be counteracted by introducing a non-reactive buffer gas \cite{Dehmelt:69,Moriwaki:92} that reduces ion temperature through collisional damping of hot ion motion. The buffer gas allows us to initially load the trap and determine the value of stray fields. After the stray fields have been compensated, the trap can be loaded at ultra-high vacuum (UHV). 

\begin{figure}[h]
\begin{center}
\includegraphics[width=7cm]{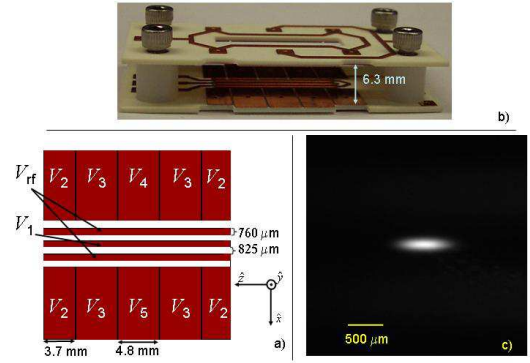}
\caption{ (color online) (a) Layout of trap electrodes, each labeled with the voltage applied; all but $V_{\rm rf}$ are dc.  The space around the long
electrodes ($V_{\rm rf}$ and $V_1$) has been milled out. Coordinates
referenced as shown define the origin at the trap center and on the
chip surface.  (b) Photograph showing the top electrode plate mounted
$6.3$ mm above the trap.  The top plate has a slit for ion fluorescence
detection; a dc voltage $V_{top}$ applied to it can deepen the trap
depth. (c) CCD image of trapped strontium ions. \vspace{-3ex}}
\label{fig:pcb}
\end{center}
\end{figure}

%%%%%%%%%%%%%%%%%%%%%%%%%%%%%%%%%%%%%%%%%%%%%%%%%%%%%%%%%%%%%%%%%%%%
% Experimental setup 
%%%%%%%%%%%%%%%%%%%%%%%%%%%%%%%%%%%%%%%%%%%%%%%%%%%%%%%%%%%%%%%%%%%%

We demonstrate this loading method using strontium ions in a
$\sim 1$ mm-scale surface electrode trap. Following the design of a
traditional four-rod linear Paul trap system \cite{Berkeland:02}, the
trap is mounted in a standard UHV chamber pumped down to $\sim
10^{-9}$ torr, loaded with $^{88}\mathrm{Sr}^+$ by electron
impact ionization of neutral atoms from a resistive oven source,
and driven by an externally mounted helical resonator.  An optional,
controlled buffer gas environment of up to $10^{-4}$ torr of
ultra-pure helium is provided though a sensitive leak valve, monitored
with a Bayard-Alpert ion gauge.

Our surface electrode ion trap has five
electrodes \cite{Chiaverini:05,Pearson:06}: one center electrode at
ground, two at rf potential, and two segmented dc
electrodes (\fig{pcb}).  The electrodes are copper, deposited on a low
rf loss substrate (Rogers 4350B), and fabricated by Hughes Circuits
following standard methods for microwave circuits.  In the loading
region, slots are milled between the rf and dc electrodes to prevent
shorting due to strontium buildup.  The inner surfaces are plated with
copper to minimize trap potential distortion due to accumulation of
stray surface charges.  The trap surface is polished to a $1$ $\mu$m
finish to reduce laser scatter into the detector.

Ions are detected by laser induced fluorescence of the main $422$ nm
$5S_{1/2} \rightarrow 5P_{1/2}$ transition of
strontium \cite{Berkeland:02}, using either an electron-multiplying CCD
camera (Princeton Instruments PhotonMax) or a photomultiplier tube
(Hamamatsu H6780-04). A laser tuned to $1092$ nm addresses the
$5P_{1/2} \rightarrow 4D_{3/2}$ transition to prevent shelving from
the $P$ state to the metastable $D$ state. The two external cavity
laser diode sources are optically locked to low finesse cavities
\cite{Hayasaka:02}.  Typical laser powers at the trap center are $1.2$
mW at $1092$ nm and $20$ - $50$ $\mu$W at $422$ nm.

\begin{figure}
\begin{center}
    \mbox{
      \scriptsize a) \subfigure{\includegraphics[width=4.0cm]{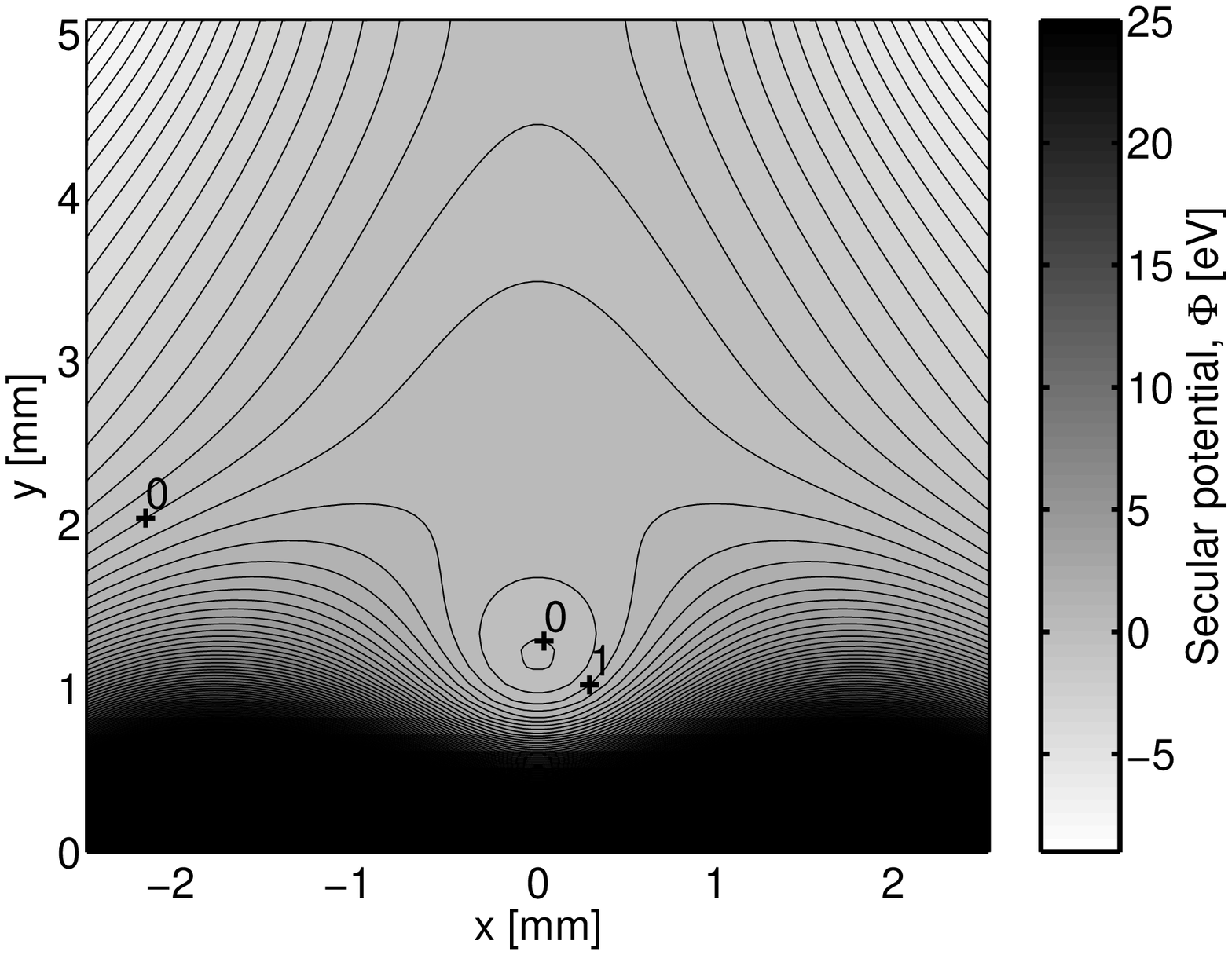}} \quad
      \scriptsize b) \subfigure{\includegraphics[width=4.0cm]{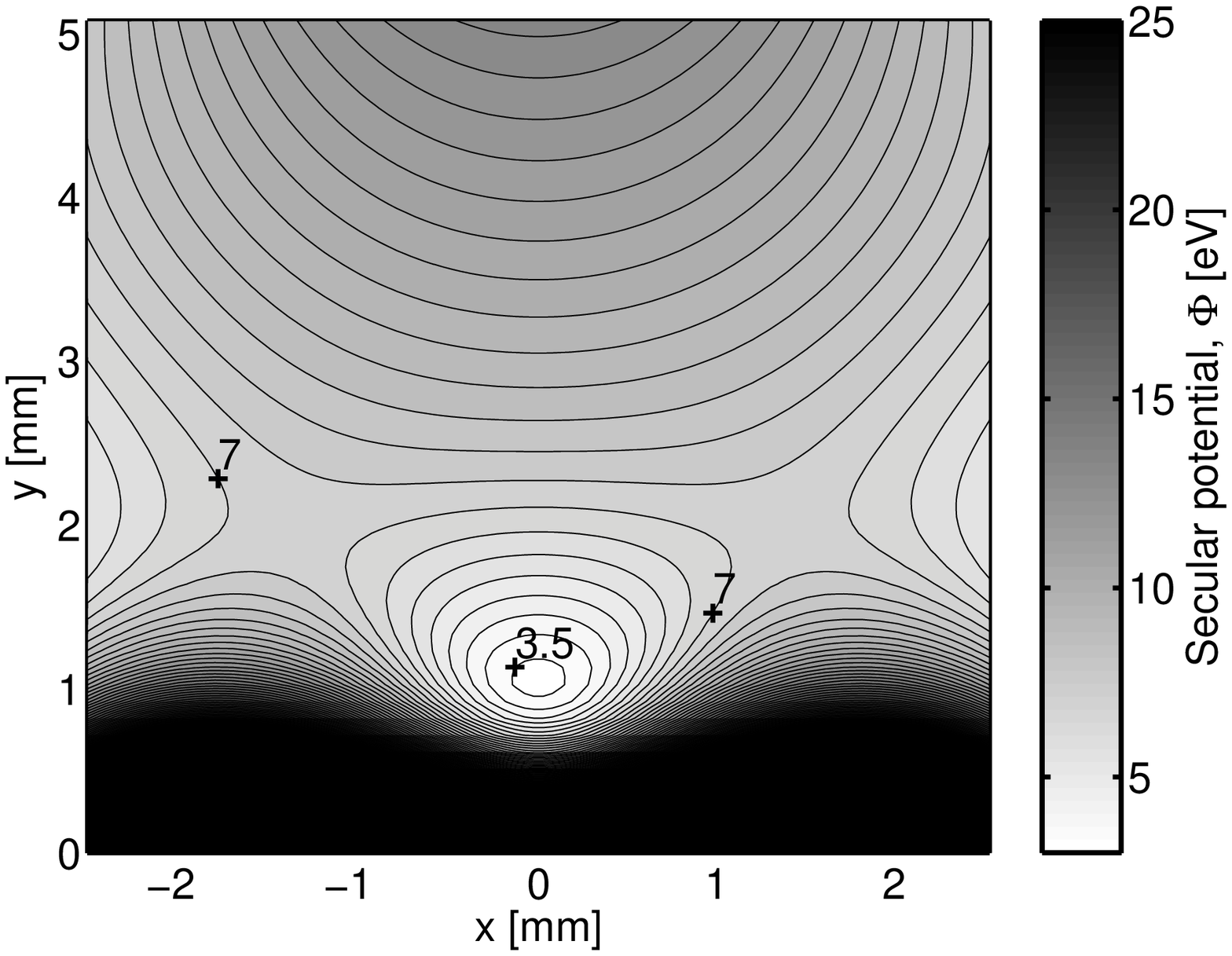}}
         }\\
    \mbox{
      \scriptsize c) \subfigure{\includegraphics[width=4.0cm]{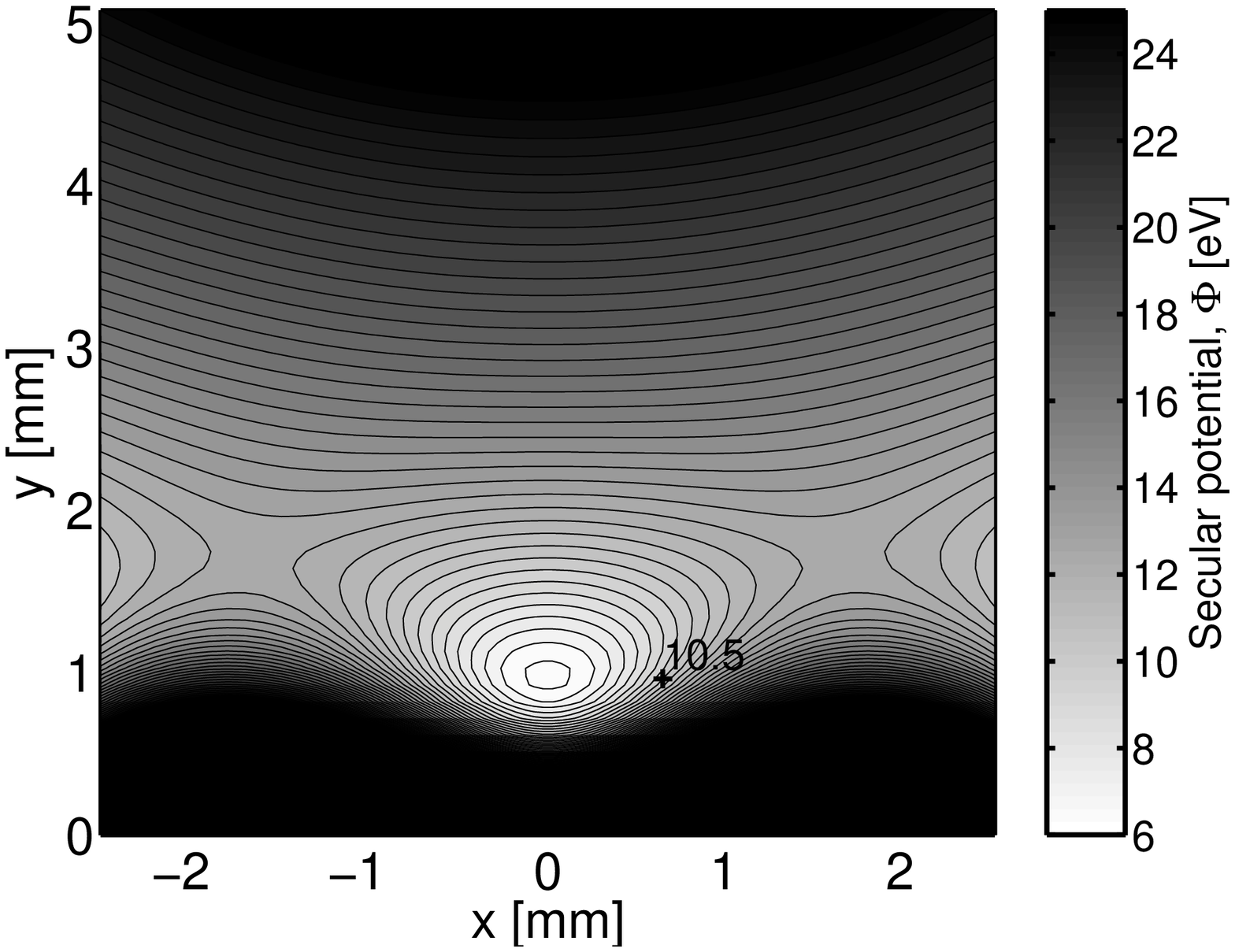}}
         }
\normalsize
\caption{Cross sections of the pseudopotential along the $\hat{x}$ and $\hat{y}$ directions, for $V_{\rm rf}= 1260$ V at $\Omega/2\pi= 7.6$ MHz, $V_2= 110$ V, and $V_3=-50$ V. The three figures a, b, and c correspond to $V_{top}$ voltages of -25.4 V, 0 V, and 15 V respectively. Micromotion compensation is expected in the -25.4 V case, but with a depth of only 1 eV, while the uncompensated 15 V case has an expected depth of 5.4 eV.}
\label{fig:pot}
\end{center}
\end{figure}

\begin{figure}
\begin{center}
\includegraphics[width=6.0cm]{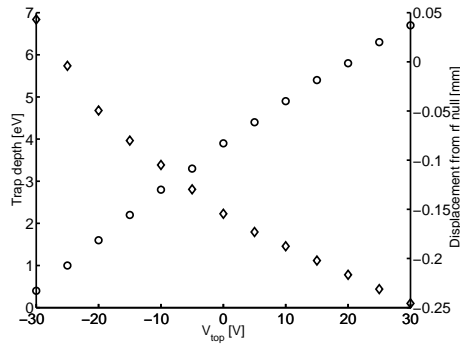}
\caption{Calculated values of trap depth (circles) and ion displacement from the rf null (diamonds) as $V_{top}$ is varied. As trap depth is increased, the displacement of the ion cloud from the rf null leads to increased micromotion.}
\label{fig:depth_pos_vtop}
\end{center}
\end{figure}

% Trap potential
The first step for loading a surface electrode trap
is determination of the ideal compensation voltages needed to offset
the inherent asymmetry.  We determine these potentials numerically
(using CPO, a boundary element electrostatic solver \cite{Brkic:06}),
by computing the rf and dc potentials ($\phi_{\rm rf}\cos{\Omega t}$ and
$\phi_{\rm dc}$), which give the secular potential $\Phi =
Q^2\l|\nabla\phi_{\rm rf}\r|^2 /4m\Omega^2+Q\phi_{\rm dc}$
% \be 
% 	\Phi = \frac{Q^2}{4m\Omega^2}\l|\nabla\phi_{\rm rf}\r|^2+Q\phi_{\rm dc}
% \,,
% \ee 
where $m$ is the ion mass and $Q$ is the ion charge. Typically,
$V_{\rm rf}$ is of $500$-$1200$ V amplitude at $\Omega/2\pi = 7.6$ MHz,
and dc electrode voltages (as defined in Fig.~\ref{fig:pcb}) are $V_4
= V_5 = 0$ V, $V_2 = 110$ V, and $V_3 = -50$ V. Shown in
\fig{pot} is a cross-section of the secular potential in the
$\hat{x}$-$\hat{y}$ plane, at $z=0$, for three different values of $V_{top}$. As demonstrated in \fig{depth_pos_vtop}, with
these voltages and $V_{top} = -25.4$ V applied to the top electrode,
the trap should be compensated with a trap depth of $1.0$ eV.
The trap depth can be increased to $5.4$ eV by setting $V_{top} = 15$
V, at the cost of increased micromotion.
 
%%%%%%%%%%%%%%%%%%%%%%%%%%%%%%%%%%%%%%%%%%%%%%%%%%%%%%%%%%%%%%%%%%%%%%%%%%%%
% Buffer gas and compensation
%%%%%%%%%%%%%%%%%%%%%%%%%%%%%%%%%%%%%%%%%%%%%%%%%%%%%%%%%%%%%%%%%%%%%%%%%%%%

\begin{figure}
\begin{center}
\includegraphics[width=7cm]{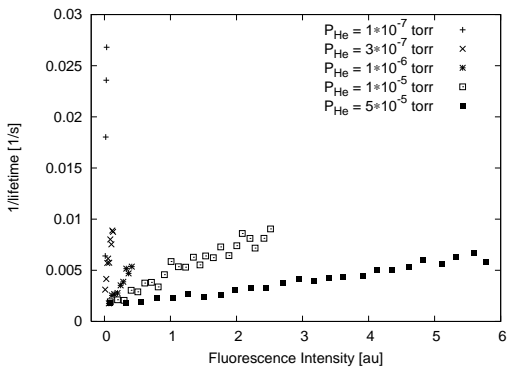}
\caption{Plot of 1/lifetime as a function of fluorescence intensity at five different buffer gas pressures in an uncompensated trap. These data show that long lifetimes can be obtained at nearly any buffer gas pressure but at very different cloud sizes as measured by fluorescence intensity. The optimum settings, long lifetimes and large clouds, are obtained at high buffer gas pressure. \vspace{-3ex}}
\label{fig:decay}
\end{center}
\end{figure}

These ideal compensation voltages often differ substantially
from actual ones, due to the presence of unknown stray charges in the
trap.  A variety of techniques have been developed to experimentally
determine the appropriate voltages, including examination of the
single ion spectrum \cite{Raab:00,Lisowski:05}, the correlation
between ion fluorescence and the rf drive phase \cite{Berkeland:98},
and the change in ion position with pseudopotential depth
\cite{Berkeland:98}. The first two methods require cold and small ion
clouds necessitating good initial compensation.  We use the last method, which is also applicable to large hot ion clouds.

We employ buffer gas cooling to load and maintain the large ion clouds
needed for experimental determination of appropriate compensation
voltages.  Initially, when the cloud center is 0.2 mm from the rf node, the size and lifetime of the loaded cloud depends strongly
on the buffer gas pressure (\fig{decay}). Notably, lifetimes at UHV were too short to be measured in the uncompensated trap.
Based on the data in \fig{decay}, we perform our compensation experiments at $1 \times 10^{-5}$
torr. This pressure yields an excellent signal to noise ratio ($\sim
200$ for a $50$ ms integration time with the photomultiplier tube) and long ion
lifetime ($\sim 300$ s) without overburdening the ion pump.

An accurate value of the stray dc field can be calculated from the
cloud motion using the following model. The electric field along a
coordinate $x$, at the rf node, is well approximated by $E(x) = E_0 +
E_1 x$.  For an rf pseudopotential with secular frequency $\omega$,
the ion motion follows $m \ddot{x} + m \omega^2 x + e E(x) = 0$, which
results in a new secular frequency $\omega_1 = \sqrt{(\omega^2 + e
E_1/m)}$, and a new cloud center position $x_0 = {e E_0}/{m
\omega_1^2}$. By measuring both the secular frequency and the ion
center, one can determine $E_0$.

We experimentally determine $E_0$ by measuring the
cloud center position as a function of applied voltages. The $1092$ nm
laser is configured to illuminate the entire trapping region, while
the $422$ nm laser is focused to a $60~\mu$m spot; the focal point is
translated in the $\hat{x}$-$\hat{y}$ plane by using a precision
motorized stage. Ion cloud fluorescence intensity, measured by the
PMT, is recorded as a function of laser position, and fit to a
Gaussian centered at the ion cloud position \cite{Neuhauser:88}.  This
measurement is then repeated at $10$ different rf voltages, and a
linear fit of the cloud center positions to $1/\omega_1^2$ determines
the stray dc field value $E_0$. $\omega_1$ is determined by applying an oscillating voltage on $V_5$ of 250 mV and observing dips in the ion fluorescence.

\begin{figure}
\begin{center}
\includegraphics[width=7cm]{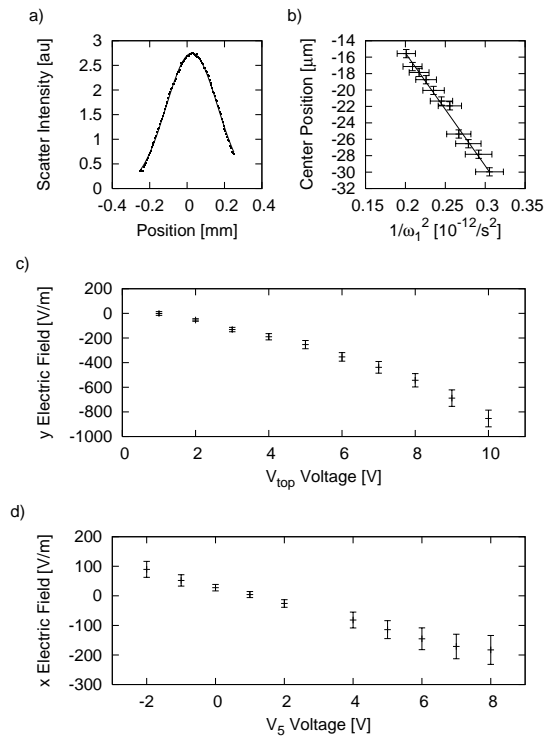}
\caption{Measurement results showing compensation of micromotion in
the trap at a buffer gas pressure of $1 \times 10^{-5}$ torr.  (a)
Cloud intensity profile along the $\hat{y}$ axis, fit to a Gaussian,
for a representative value of the the $\hat{y}$ compensation voltage,
$V_{top}$.  (b) Linear fit of the cloud center position versus
$1/\omega_1^2$ where $\omega_1$ is the secular frequency of the ion
motion, yielding the electric field along $\hat{y}$ at the rf node.
(c) Plot of the $\hat{y}$ electric field as a function of the $V_{top}$
compensating voltage, showing that the stray field is minimized at
$V_{top} = 1.0$ V. (d) Plot of the $\hat{x}$ electric field as a function of the
middle electrode voltage $V_5$, showing compensation at $V_{5} = 1.3$
V. \vspace{-3ex} }
\label{fig:exptfig}
\end{center}
\end{figure}

The data obtained, shown in \fig{exptfig}, give an excellent match of
the cloud intensity to a Gaussian fit, allowing measurement of the
cloud center to within $\pm 0.5~\mu$m. Thus, the measurement of stray
fields is precise to about $\pm 10$V/m at zero stray field. From the
stray field measurements, we determine the required compensation
voltages to be $V_{top} = 1.0 \pm 0.1$ V and $V_5 = 1.3 \pm 0.3$
V. The estimated residual displacement of a single ion at these
voltages is less than $0.2~\mu$m.  The nonlinear dependence of the dc
electric field along $\hat{y}$ on the top electrode voltage is due to
the strong anharmonicity of the trap in the vertical direction,
unaccounted for in the simple linear model employed in the analysis.

%%%%%%%%%%%%%%%%%%%%%%%%%%%%%%%%%%%%%%%%%%%%%%%%%%%%%%%%%%%%%%%%%%%%%%%%%%%%%
% Conclusions
%%%%%%%%%%%%%%%%%%%%%%%%%%%%%%%%%%%%%%%%%%%%%%%%%%%%%%%%%%%%%%%%%%%%%%%%%%%%%

The difference between measured and ideal compensation voltages is
evidence of anisotropic stray fields, caused by undetermined surface
charges.  The estimated stray fields along $\hat{x}$ are comparable to
those reported for 3D traps \cite{Berkeland:98}. However, the stray
fields along $\hat{y}$ are $10$ times larger. The $26$ V difference
between the calculated and measured values of $V_{top}$ at
compensation suggests significant electron charging on either the trap
surface, the top plate, or the top observation window.

In summary, we have loaded a surface electrode ion
trap by electron impact ionization at UHV by using an uncompensated trap to increase trap depth and a large cloud in buffer gas to find the compensation values. The results suggest that the open geometry of the trap makes it more susceptible to stray surface charges. The technique demonstrated
will likely be useful for the loading of complex and integrated surface electrode ion traps.

Support for this project was provided in part by the JST/CREST Urabe
Project, and MURI project F49620-03-1-0420.  We thank Rainer Blatt,
Richart Slusher, Vladan Vuletic, and David Wineland for helpful
discussions.

%%%%%%%%%%%%%%%%%%%%%%%%%%%%%%%%%%%%%%%%%%%%%%%%%%%%%%%%%%%%%%%%%%%%%%%%%%%%%
\bibliographystyle{apsrev}

%%%%%%%%%%%%%%%%%%%%%%%%%%%%%%%%%%%%%%%%%%%%%%%%%%%%%%%%%%%%%%%%%%%%%%%%%%%%%
%%%%%%%%%%%%%%%%%%%%%%%%%%%%%%%%%%%%%%%%%%%%%%%%%%%%%%%%%%%%%%%%%%%%%%%%%%%%%
\end{document}